\documentclass{elsart}
\usepackage{epsfig}
\usepackage{amssymb}
\usepackage{rotating}

\begin{document}

%
\begin{frontmatter}

\title{First limits on WIMP nuclear recoil signals in ZEPLIN-II: a two phase xenon detector for dark matter detection}

\author[RAL]{G. J. Alner},
\author[ICL,RAL]{H. M. Ara\'ujo},
\author[ICL]{A. Bewick},
\author[RAL,ICL]{C. Bungau},
\author[RAL]{B. Camanzi},
\author[SHE]{M. J. Carson},
\author[OXF]{R. J. Cashmore},
\author[SHE]{H. Chagani},
\author[LIP]{V. Chepel},
\author[UCL]{D. Cline},
\author[ICL]{D. Davidge},
\author[SHE]{J. C. Davies},
\author[SHE]{E. Daw},
\author[ICL]{J. Dawson},
\author[RAL]{T. Durkin},
\author[RAL,ICL]{B. Edwards},
\author[SHE]{T. Gamble},
\author[TEX]{J. Gao},
\author[EDI]{C. Ghag},
\author[ICL]{A. S. Howard},
\author[ICL]{W. G. Jones},
\author[ICL]{M. Joshi},
\author[EDI]{E. V. Korolkova},
\author[SHE]{V. A. Kudryavtsev},
\author[SHE]{T. Lawson},
\author[ICL]{V. N. Lebedenko},
\author[RAL]{J. D. Lewin},
\author[SHE]{P. Lightfoot},
\author[LIP]{A. Lindote},
\author[ICL]{I. Liubarsky},
\author[LIP]{M. I. Lopes},
\author[RAL]{R. L\"{u}scher},
\author[SHE]{P. Majewski},
\author[SHE]{K Mavrokoridis},
\author[SHE]{J. E. McMillan},
\author[SHE]{B. Morgan},
\author[SHE]{D. Muna},
\author[EDI]{A. St.J. Murphy},
\author[LIP]{F. Neves},
\author[SHE]{G. G. Nicklin},
\author[UCL]{W. Ooi},
\author[SHE]{S. M. Paling},
\author[LIP]{J. Pinto da Cunha},
\author[EDI]{S. J. S. Plank},
\author[RAL]{R. M. Preece},
\author[ICL]{J. J. Quenby},
\author[SHE]{M. Robinson},
\author[UCL,PIS]{F. Sergiampietri},
\author[LIP]{C. Silva},
\author[LIP]{V. N. Solovov},
\author[RAL]{N. J. T. Smith\corauthref{cor1}},
\corauth[cor1]{Corresponding author; address: Particle Physics Dept., CCLRC Rutherford
Appleton Laboratory, Chilton, Didcot, Oxon, OX1 0QX, UK}
\ead{n.j.t.smith@rl.ac.uk}
\author[RAL,UCL]{P. F. Smith},
\author[SHE]{N. J. C. Spooner},
\author[ICL]{T. J. Sumner},
\author[ICL]{C. Thorne},
\author[SHE]{D. R. Tovey},
\author[SHE]{E. Tziaferi},
\author[ICL]{R. J. Walker},
\author[UCL]{H. Wang\corauthref{cor2}},
\corauth[cor2]{Corresponding author; address: Department of Physics \& Astronomy, University of California, Los Angeles, USA}
\ead{wangh@physics.ucla.edu}
\author[TEX]{J. White} \&
\author[ROC]{F. L. H. Wolfs}

\address[RAL] {Particle Physics Dept., CCLRC Rutherford Appleton Laboratory, UK}
\address[ICL] {Blackett Laboratory, Imperial College London, UK}
\address[SHE] {Department of Physics \& Astronomy, University of Sheffield, UK}
\address[OXF] {Brasenose College and Department of Physics, University of Oxford, UK}
\address[LIP] {LIP--Coimbra \& Department of Physics of the University of Coimbra, Portugal}
\address[UCL] {Department of Physics \& Astronomy, University of California, Los Angeles, USA}
\address[EDI] {School of Physics, University of Edinburgh, UK}
\address[TEX] {Department of Physics, Texas A\&M University, USA}
\address[PIS] {INFN Pisa, Italy}
\address[ROC] {Department of Physics and Astronomy, University of Rochester, New York, USA}

\newpage
\begin{abstract}
Results are presented from the first underground data run of ZEPLIN-II, a 31~kg two phase xenon detector developed to observe nuclear recoils from hypothetical weakly interacting massive dark matter particles. Discrimination between nuclear recoils and background electron recoils is afforded by recording both the scintillation and ionisation signals generated within the liquid xenon, with the ratio of these signals being different for the two classes of event. This ratio is calibrated for different incident species using an AmBe neutron source and $^{60}$Co $\gamma$-ray sources. From our first 31 live days of running ZEPLIN-II, the total exposure following the application of fiducial and stability cuts was 225~kg$\times$days. A background population of radon progeny events was observed in this run, arising from radon emission in the gas purification getters, due to radon daughter ion decays on the surfaces of the walls of the chamber. An acceptance window, defined by the neutron calibration data, of 50\% nuclear recoil acceptance between 5~keV$_{ee}$ and 20~keV$_{ee}$, had an observed count of 29 events, with a summed expectation of 28.6$\pm$4.3 $\gamma$-ray and radon progeny induced background events. These figures provide a 90\% c.l. upper limit to the number of nuclear recoils of 10.4 events in this acceptance window, which converts to a WIMP-nucleon spin-independent cross-section with a minimum of  $6.6\times10^{-7}$~pb following the inclusion of an energy dependent, calibrated, efficiency. A second run is currently underway in which the radon progeny will be eliminated, thereby removing the background population, with a projected sensitivity of $2\times10^{-7}$~pb for similar exposures as the first run.
\end{abstract}


\end{frontmatter}

%
%
\section{Introduction}
Several underground experiments are in operation or under development throughout the world to search for the low energy nuclear recoils that would result from elastic collisions between the hypothetical Galactic weakly interacting massive dark matter particles (WIMPs) and the nuclei of normal matter \cite{dm01,dm02,dm03,dm04}. A key feature of such experiments is that they require some means of discriminating nuclear recoils from the much larger number of electron recoils that will in general be present from background $\gamma$-ray interactions or $\beta$-decay events. ZEPLIN-II \cite{zii_a,zii_b} is a two-phase (liquid/gas) xenon detector constructed by the ZEPLIN II Collaboration\footnote{University of Edinburgh, Imperial College London, LIP-Coimbra, University of Rochester, CCLRC Rutherford Appleton Laboratory, University of Sheffield, Texas A\&M University, UCLA.} as  part of a long term development programme of liquid xenon dark matter detectors \cite{zeplin1,zeplin2,zeplin3,zeplin4}. ZEPLIN-II is operated at the Boulby underground laboratory in the U.K., with the aim of observing these low energy elastic nuclear recoils due to WIMPs. 

A liquid xenon target is afforded discrimination power between incident species by the fact that particle interactions will produce both VUV scintillation light and ionisation (electrons), in a ratio which differs for nuclear and electron recoils \cite{kubota79,hitachi83,davies94}.  An important implementation of this is the use of a two-phase system \cite{dolgoshein70,cline00} in which two signals are produced for each event: from the primary scintillation light (S1); and from the use of electric fields to drift the charge to the liquid surface, from where it is extracted into a high E-field gas region to produce a second electroluminescence pulse (S2).

Event-by-event discrimination is possible by comparing the S2/S1 ratio for each interaction within the liquid xenon with calibrated signals from neutron and $\gamma$-ray sources, providing the required nuclear and electron recoils. In addition to good incident species discrimination, a direct dark matter detector requires a low intrinsic background rate to observe the rare WIMP interactions. This is achieved through the use of radio-pure materials in construction, external $\gamma$-ray and neutron shielding, active veto systems and deep underground operation to remove cosmic ray induced backgrounds. ZEPLIN-II is primarily constructed from low background materials, operated 1070~m underground, surrounded by an active liquid scintillator veto and passive lead and hydrocarbon shielding. The kinematics of the WIMP interaction on the target nuclei produces a featureless and soft ($\lesssim$100~keV) recoil energy spectrum, requiring detectors with low energy thresholds. The ZEPLIN-II detector has sufficient sensitivity to the scintillation light to provide a usable electron recoil equivalent energy (keV$_{ee}$) threshold of 5~keV$_{ee}$ and sensitivity to single electrons extracted from the liquid surface.

We report here results from the first underground science run, of 31 days livetime. The target mass of ZEPLIN-II is 31~kg, with a fiducial mass of 7.2~kg once all spatial selection cuts are applied. In this 225~kg$\times$days exposure run, 29 events are seen in an acceptance window defined between of 5~keV$_{ee}$ and 20~keV$_{ee}$, with a 50\% nuclear recoil acceptance efficiency. These observed events arise from two sources: a small number of expected $\gamma$-ray induced electron recoils; and an unexpected population of events due to recoiling radon daughters on the polytetrafluoroethylene (PTFE) surfaces of the detector. Expectation calculations for these two populations yield a prediction of 28.6$\pm$4.3 events in total. This leads to a 90\% c.l. upper limit of 10.4 events for nuclear recoils within this acceptance window which, allowing for trigger and selection efficiencies and the detector response, provides a WIMP-nucleon cross-section limit which reaches a minimum of  $6.6\times10^{-7}$~pb at a WIMP mass of 65~GeV.

%
\section{The ZEPLIN-II Detector}
A detailed description of the ZEPLIN-II detector will be given in a companion instrument paper, including cryogenic and gas systems, and operational details. The data acquisition system and data reduction procedures are also described elsewhere \cite{alner07}. This paper includes only those details relevant to the calculation of a dark matter limit.

\subsection{Liquid xenon detector principles}
Nuclear recoil discrimination in liquid xenon arises from measuring both scintillation light and ionisation produced during an interaction. The energy deposited appears in different channels which, with the exception of a phonon component, involve radiative processes:

\begin{itemize}
\item{The production and radiative decay of excited Xe$_2^*$ states. Decay of the singlet and triplet states of the Xe$_2^*$ excimer to the ground state results in emission of 175 nm photons, with characteristic decay times of 3 ns and 27 ns respectively, being followed by the dissociation of the Xe$_2$ molecule. The light yield for liquid xenon at zero electric field is $\sim$30-80~photons per keV of deposited energy for $\gamma$-rays \cite{doke99,chepel05b}. The energy loss rate ($dE/dx$) of a particle determines the proportion of energy channelled into these states as well as the singlet/triplet ratio \cite{kubota79,akimov02}. As a result, nuclear recoils produce scintillation pulses which are significantly faster than those from electron recoils}
\item{The recombination of ionised Xe$_2^+$ states. The ionised Xe$_2^+$ dimer can recombine with electrons along the particle track to produce Xe$_2^*$ excimers, which decay as above.  When radiative recombination is allowed to occur then the $dE/dx$ of the particle determines the recombination time; this is extremely fast for nuclear recoils ($<$1~ns), but much slower for electron recoils ($\sim$40~ns).  This discrimination principle was utilised in the zero field detector, ZEPLIN-I \cite{zeplin1}.}
\end{itemize}

For two-phase detectors, such as ZEPLIN-II, the presence of an electric field allows the ionisation to be collected and measured indirectly through electroluminescence caused in the gas phase. This is made possible by the ease with which electrons can be drifted through the liquid phase and extracted into the gas phase; the efficiency of the ionisation separation process depends on the initial linear ionisation density. This extraction will be most pronounced for $\gamma$-ray interactions where the ionisation track tends to be less dense and the electric field will be more effective at separating the free electrons from the ions. Once separated, the electrons will drift in the direction defined by the applied electric field until they reach the xenon liquid surface where they can be extracted into the gas phase; here a higher field region causes the extracted charge to produce secondary scintillation, or electroluminescence, which is proportional to the amount of charge extracted \cite{conde77}. 

Thus, in ZEPLIN-II both the primary scintillation and the ionisation signals are seen as vacuum ultra-violet (VUV) light pulses, with a time delay between them: the primary scintillation signal occurs first with the second signal (secondary scintillation caused by the electrons accelerated in the gas) occuring after charge drift and extraction from the liquid. For a given number of primary photons, a $\gamma$-ray will produce many more secondary photons than an $\alpha$-particle or a nuclear recoil. This is the basis for the discrimination power of two-phase detectors. Tecnically, the ratio between integrated areas of the two signals was measured.

On an event by event basis the scintillation and ionisation signals are anti-correlated \cite{conti03}. This arises as the recombination of the ionisation contributes to the overall primary scintillation signal, commensurately decreasing the ionisation signal. Although this may be used to generate a tighter energy distribution for the interactions, we observed that it did not improve our event by event discrimination between nuclear and electron recoils in S2/S1.

\subsection{ZEPLIN-II design and layout}
\begin{figure}[p]
   \centerline{\epsfig{file=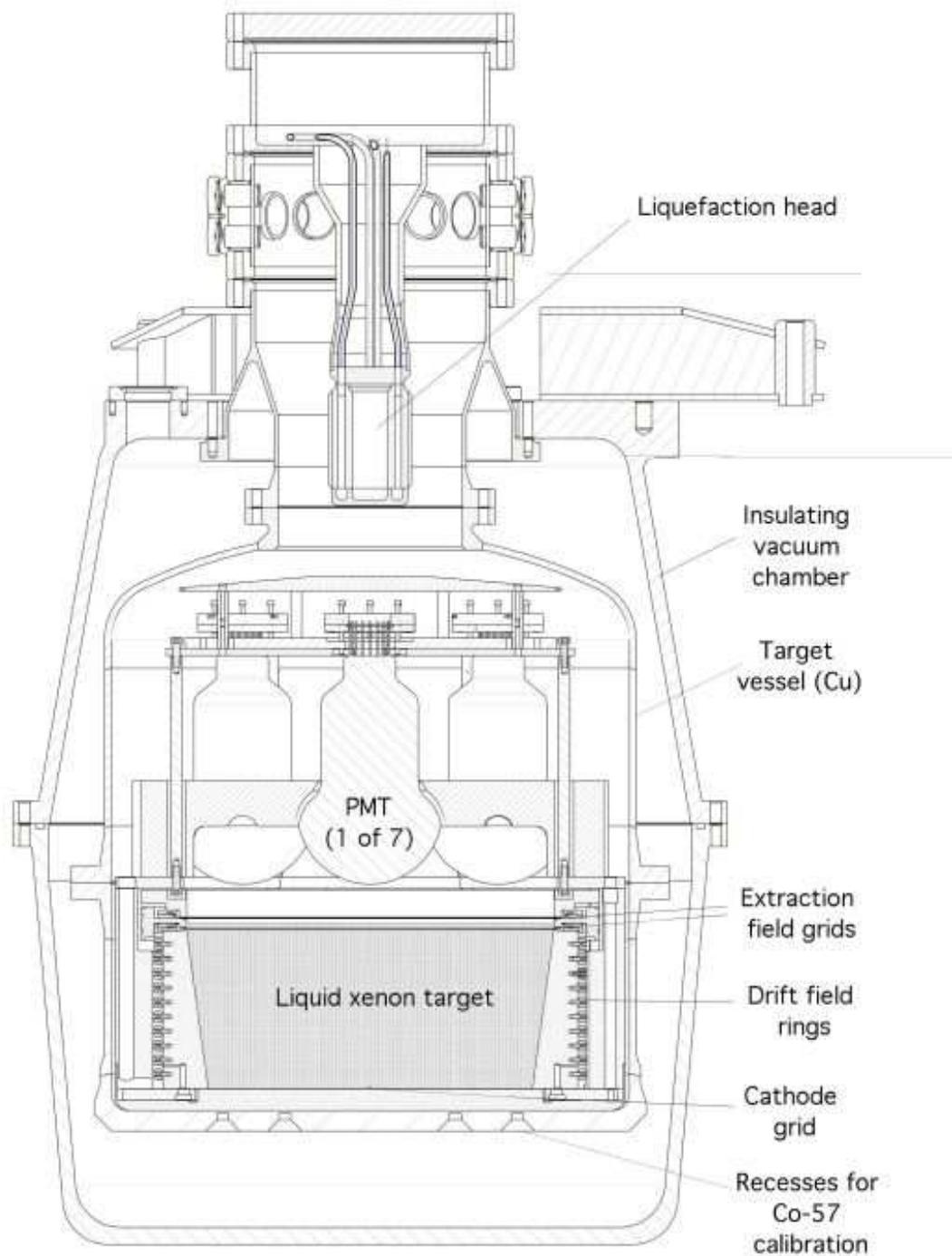,width=15.cm}}
    \caption{Schematic of the ZEPLIN-II detector. The liquid xenon volume is shown, viewed from above by 7 quartz-window photomultipliers. The electrode arrangement defines a drift region between the cathode grid and the lower extraction grid where the field is parallel and uniform (this is obtained with the help of lateral field-shaping rings embedded in the PTFE walls). The extraction region (where electroluminescence is generated) is defined by the two grids located either side of the liquid surface. Xenon liquefaction occurs on the liquefaction head, with liquid dripping onto a copper shield which deflects it away from the photomultiplier array and the active volume.}
  \label{detector}
\end{figure}

\begin{figure}[p]
   \centerline{\epsfig{file=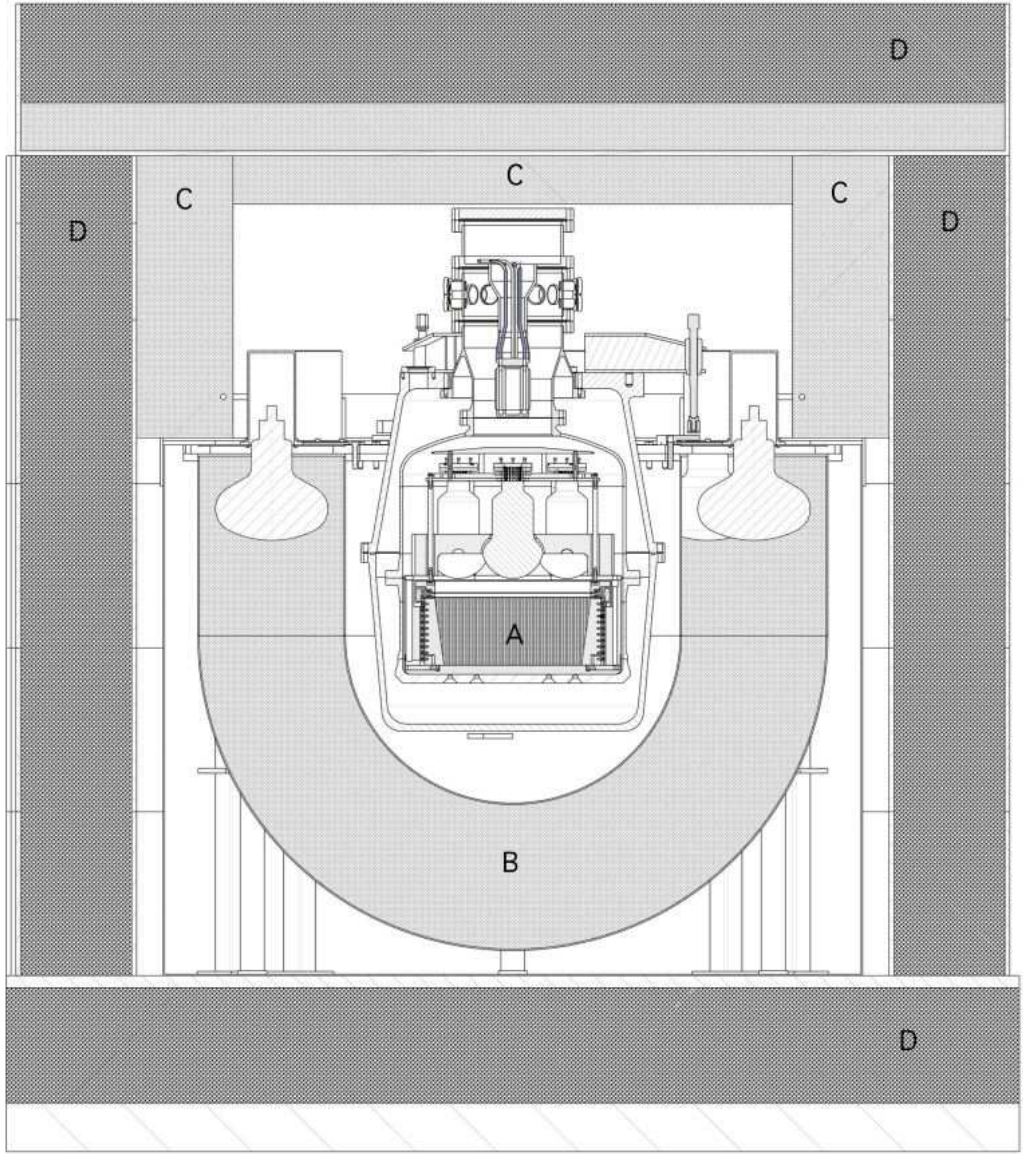,width=15.cm}}
    \caption{Arrangement of the ZEPLIN-II detector within the $\gamma$-ray and neutron shielding. The detector (A) is located in a 30~cm thick, 1~tonne liquid scintillator veto (B), with 30cm of Gd-loaded polypropylene hydrocarbon on the top surfaces (C). Surrounding the hydrocarbon shielding is a minimum of 25cm Pb $\gamma$-ray shielding (D).}
  \label{assembly}
\end{figure}

The general layout of the ZEPLIN-II detector itself is shown in Fig.~\ref{detector}. Fig.~\ref{assembly} shows the detector within its liquid scintillator veto/neutron shield and lead $\gamma$-ray shield. The target mass of 31~kg liquid xenon is viewed from above by 7 quartz-window 13~cm diameter ETL low background D742QKFLB photomultipliers \cite{etl} arranged in a hexagonal pattern inside the rolled copper target vessel. The photomultipliers have a Pt underlay plated beneath the photocathode to allow for cryogenic operations, which reduces the quantum efficiency of the photomultipliers to 17\% for 175~nm light, at room temperature. The target vessel is surrounded by a vacuum vessel of cast stainless steel alloy.   Feed-throughs for high voltage and environmental monitoring, and xenon gas and cryogenic connections emerge through the top of the vessels and pass through the shielding systems. 

The 14~cm deep liquid xenon active volume is defined by a thick PTFE ring which acts as a reflector for the VUV 175~nm scintillation light, provides a support structure for the field shaping rings and ensures a uniform electric field within the drift volume.  To collect charge from ionisation in the liquid, an electric field of 1~kV/cm is maintained through the target volume by means of a cathode mesh at the base of the target vessel and a second grid inside the liquid close to the liquid surface, parallel to the cathode mesh, with field shaping rings to achieve uniformity.  A third grid is placed above the liquid surface, to provide a strong electric field (4.2~kV/cm in the liquid and 8.4~kV/cm in the gas) for extraction of electrons from the liquid and to provide the electroluminescence region in the gas phase. This field provides a 90\% extraction efficiency of electrons from the liquid surface \cite{gushchin79} with a measured secondary yield of $\sim$230 electroluminescence photons per extracted electron from the liquid surface, at the mean operating pressure of 1.5~bar. The PTFE walls within the vessel are tapered, to minimise charge trapping, thereby allowing all charge to be drifted to the liquid surface.

The xenon gas is cooled by an IGC PFC330 Polycold refrigerator \cite{polycold} connected to a copper liquefaction head within the target chamber. The target chamber and internal structures are cooled convectively and by xenon `rain' from this liquefaction head. In operation the xenon is constantly recirculated by drawing liquid from outside the active volume using an internal heater and a Tokyo Garasu Kikai MX-808-ST diaphragm pump \cite{pump}, distilling through a SAES getter PS11-MC500 purifier \cite{saes}. A flow rate of 3~slpm was maintained to ensure sufficient purity of xenon ($>$100~$\mu$s) to allow charge collection throughout the active volume. 

\begin{figure}[p]
   \centerline{\epsfig{file=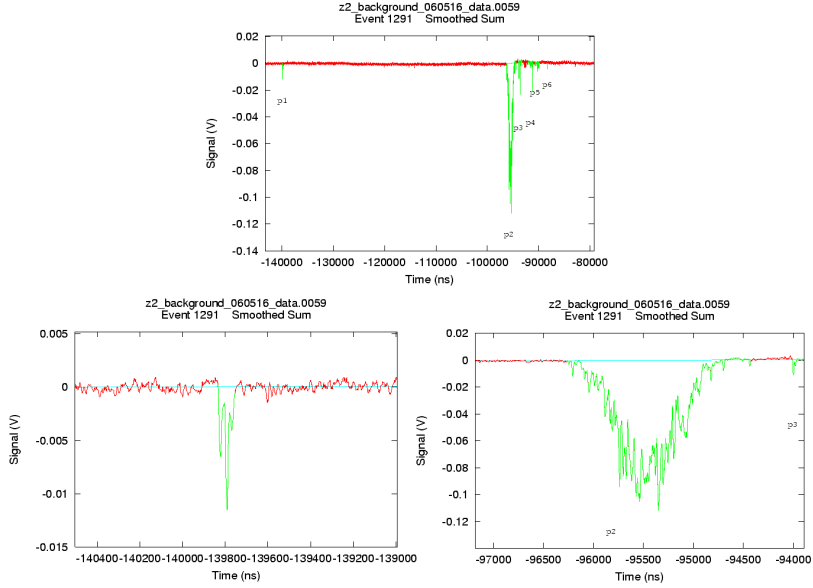,width=11.cm}}
    \caption{Typical $\gamma$-ray event recorded during the science data run, with an energy of 16~keV$_{ee}$. The upper plot shows the overall digitisation trace showing the S1 signal (labelled p1) and the S2 signal (labelled p2). The lower plots show extended traces of the S1 (left) and S2 (right) signals. The S2 signal area, which is proportional to the number of detected VUV photons, is $\sim$300 times that of the S1 signal.}
  \label{pulse}
\end{figure}

\begin{figure}[p]
   \centerline{\epsfig{file=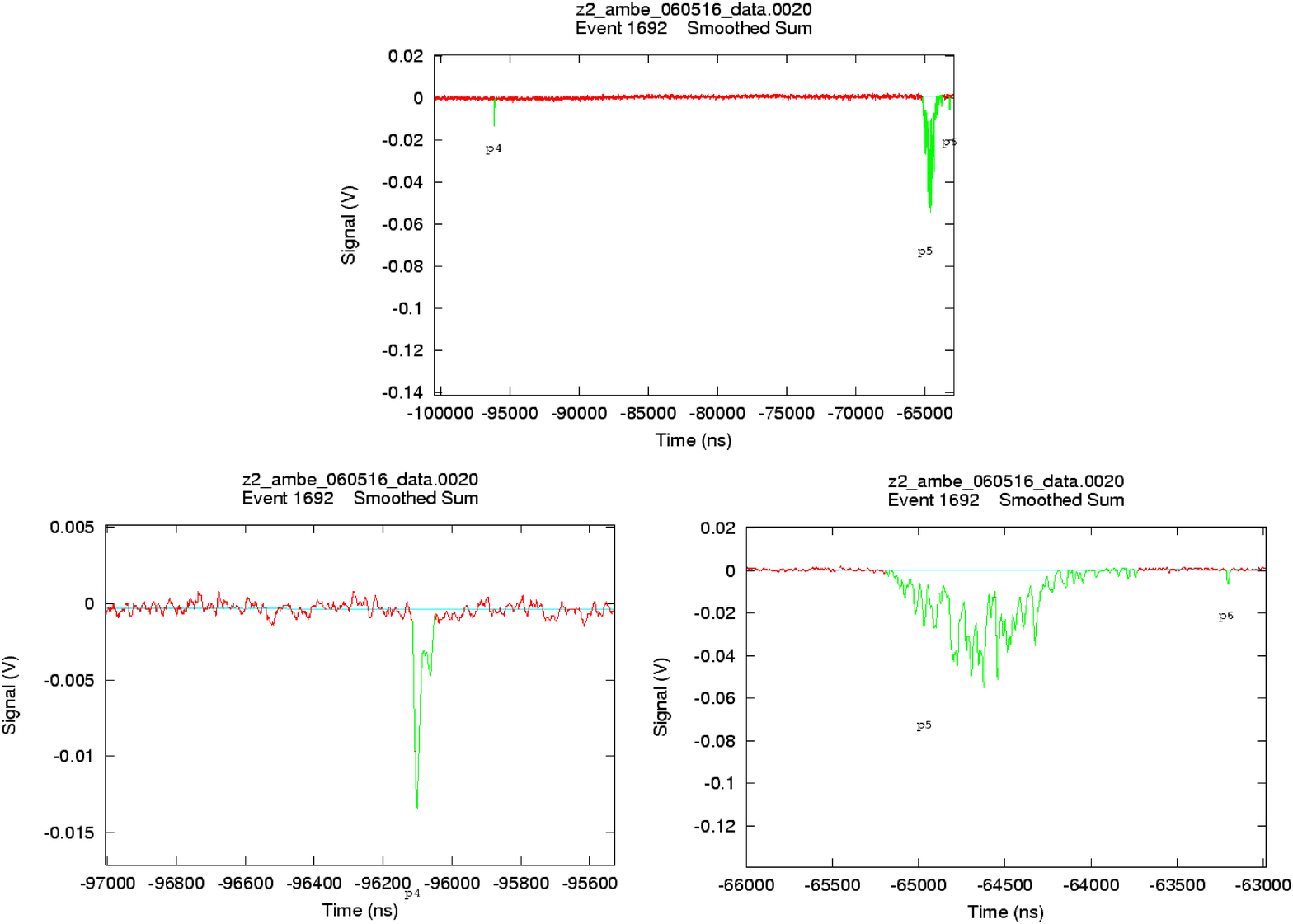,width=11.cm}}
    \caption{Example of a single scattered neutron event within ZEPLIN-II from an AmBe calibration run, with an energy of 16~keV$_{ee}$. The upper plot shows the overall digitisation trace showing the S1 signal (labelled p4) and the S2 signal (labelled p5). The lower plots show extended traces of the S1 (left) and S2 (right) signals. For neutron events the S2 signal area is $\sim$100 times that of the S1 pulse. The vertical scales are identical to Fig. \ref{pulse}.}
  \label{n_pulse}
\end{figure}

As a result of this arrangement, the photomultipliers register two pulses from each particle interaction in the active xenon. The first is from the direct S1 scintillation light, the second from the S2 electroluminescence signal.  The delay between the two corresponds to the drift time of the charge and is thus depth-dependent, being 73~$\mu$s for the full 14~cm depth.  Fig.~\ref{pulse} shows an example of a $\gamma$-ray interaction within the active volume, showing the typical structure of the primary and secondary signals.  Fig.~\ref{n_pulse} shows an example of the S1 and S2 signals for a nuclear recoil event (from neutron scattering) illustrating the typically lower value of S2/S1 for the latter. 

\subsection{Data acquisition and trigger}
Full details of the data acquisition system and data reduction techniques are given in a companion paper \cite{alner07}, but are summarised here. The signals from the 7 photomultipliers are split passively through Suhner 4901.01.A 2~GHz 50~$\Omega$ power dividers\cite{suhner}, with one signal being used to create the trigger, the second being digitised as the event waveform. The photomultiplier signals are digitised with 8 bit resolution at 500~Msamples/s, with a 150~MHz bandwidth and a depth of up to 2~Msamples/channel. This digitisation is performed using cPCI based DC265 Acqiris\cite{Acqiris} digitisers within a CC103 Acqiris crate, under control of a Linux based PC. The last channel of the 8 channel digitisation system is used to digitise the summed output from the liquid scintillator veto.

The second arm of the split signals are amplified ($\times$10), discriminated at 2/5 photoelectron and fed to a majority logic trigger which is set to fire when 5 photomultipliers out of the 7 see a signal above threshold. A high level inhibit signal is also applied based on the output from the central photomultiplier in the array to minimise the DAQ deadtime, where events which would heavily saturate the digitisers are vetoed in the trigger hardware. This trigger philosophy is based on the ability to trigger on the secondary electroluminescence signal for low energy events, where the electroluminescence signal is distributed across the majority of the photomultipliers. For high energy events the trigger will occur on the commensurately larger primary scintillation pulse. Accordingly, the digitisers are set to acquire data 100~$\mu$s before and after the trigger point, allowing a `look-back' for primaries when the secondary triggers and ensuring the full depth of the xenon volume will be covered within the pulse traces, whether the trigger is on the primary or secondary. 

To minimise accidental coincidences between single photoelectrons, a software selection cut is applied that requires a three fold (at 2/5 photoelectron) coincidence in the primary signal. A software cut is also made to eliminate multiple scattering events, for example neutron double and triple scattering, which are of no relevance to the experiment or calibration.  Events for which S2 saturates are also rejected by the analysis, since these are all $\gamma$-ray or $\alpha$-particle events, the photomultiplier gain being adjusted to ensure that events in the nuclear recoil region do not produce saturation. This trigger philosophy avoids the inefficiency at small photoelectron numbers for a simple primary trigger because the larger secondary pulse will trigger 5 photomultipliers with $>$99\% efficiency, and the look-back technique finds 3 fold primaries that would not trigger the electronics. The efficiciencies associated with the hardware and software trigger, and the DAQ saturation are detailed in $\S$\ref{efficiencies}.

\subsection{Background studies and shielding/veto systems}
The ZEPLIN-II experiment is located in the Boulby salt and potash mine (Cleveland, UK) at a vertical depth of 1070~m (2805~m water-equivalent shielding), reducing the cosmic ray muon flux by a factor of about $10^{6}$ to a level of (4.09$\pm$0.15)$\times10^{-8}$ ~muons/cm$^2$/s \cite{robinson03}. The average radioactive contamination of the salt rock is 65~ppb U, 130~ppb Th, and 1100~ppm K \cite{smith04}.  To attenuate both $\gamma$-ray and neutron background from both radioactivity and residual muons, the detector is surrounded by an outer 25~cm Pb $\gamma$-ray shield and an inner 30~cm hydrocarbon neutron shield, the latter consisting of a vessel of liquid scintillator and a roof of solid hydrocarbon blocks (Fig.~\ref{assembly}), both with Gd-layering. This shielding system is the same as that used previously for the single phase ZEPLIN-I experiment \cite{zeplin1}.

The most important intrinsic background for nuclear recoils from WIMP collisions is that of nuclear recoils from neutron backgrounds.  The latter can arise from cosmic ray muon spallation reactions and secondary cascades, and contamination of surroundings or detector components with  uranium and thorium through spontaneous fission (mainly of $^{238}$U) and the ($\alpha$, n) reaction.   For the present experiment, the various sources of neutron background have been estimated by detailed simulations\cite{carson04,araujo05b,bungau05}, for various site depths including that of the Boulby Mine.

In the nuclear recoil range 25-50~keV, the expected single scattering neutron event rates within ZEPLIN-II for 30~kg xenon are
\begin{enumerate}
\item $\lesssim$3 events/year from muons hitting rock, shielding and detector vessels.
\item $\sim$3 events/year from U/Th radioactivity in rock, shielding and detector vessels 
\item $\sim$3-10 events/year from U/Th in the vacuum vessel
\item $\sim$10 events/year from U/Th in the photomultiplier array
\end{enumerate}

Allowing a possible factor 3 higher for (1) and (2) from shielding gaps due to pipe routes, this gives a total estimated neutron background $<$40 events/year for single scattered events in the relevant energy range for dark matter searches. This converts to  $\ll$0.01 events/kg/day, corresponding to a WIMP-nucleon cross-section limit ~ $\ll10^{-7}$~pb.  Thus, from these prior simulations, it was concluded that ZEPLIN-II should be able to reach an order of magnitude below currently achieved sensitivities before being limited by neutron backgrounds.  

Significant rejection of this neutron background is possible using coincidence with signals from the liquid scintillator veto, since the majority of neutrons (eg from the photomultipliers) scattering in the liquid xenon will then pass into or through the liquid scintillator veto, producing a signal either by nuclear scattering or absorption on the hydrogen. The veto also records signals from cosmic-ray muons contributing to the rejection of the muon-induced background. The liquid scintillator is observed by ten 20~cm diameter photomultipliers, giving an overall measured energy threshold of about 100~keV.   Simulations indicate that up to 60\% of the neutron events in the xenon could be vetoed in this way \cite{bungau05,carson05}.   Although there is no liquid scintillator directly above the detector, the solid hydrocarbon blocks were loaded with 0.2\% Gd on average, to capture neutrons thermalised by the hydrocarbon, releasing 8~MeV in $\gamma$-rays, some fraction of which can be detected in the liquid scintillator.  The overall efficiency for vetoing low energy neutrons has been measured during AmBe neutron source calibrations and is found to be 49\%, in agreement with the above simulations \cite{daw06}. The liquid scintillator veto also enables rejection of Compton-scattered $\gamma$-ray events, measured during science data runs with a 14\% veto efficiency for $\gamma$-rays with $<$50~keV energy deposition in the active xenon volume.

Another potential source of background is from $\alpha$-particles emitted by uranium or thorium decays within the detector materials, or from radon emitted locally by uranium.  Radon daughter products will migrate to the grids or the surfaces around the active volume, through the production of positively charged ions from $\beta$-decays. Although $\alpha$-particles are emitted at MeV energies, small energy deposits  down to the keV range can occur by partial energy loss at boundary walls or close to grid wires, and these may also mimic nuclear recoils.  The most prevalent sources of these events are from the cathode and field grids. These can be rejected by a timing cut, which effectively rejects events from $\sim$1~cm of liquid at the bottom and top of the active volume and reduces the active volume by $\sim$16\% to 26~kg. More challenging are the nuclear recoils produced by the radon progeny electrostatically attracted to the side walls, since these may mimic low energy xenon recoils. To remove these events, and low energy electron recoil events from the walls, a radial cut is required based on the relative secondary signal size in each photomultiplier. This radial cut reduces the target fiducial mass by $\sim$70\% to 7.2~kg.

Although low-Kr xenon is used in this experiment,  a background of $\beta$-decays from $^{85}$Kr is expected.  From the viewpoint of this experiment these simply add to the electron recoil population from $\gamma$-ray background.  We have shown previously that  $\beta$-decays do, as expected, give scintillation pulses closely similar to those for electron recoils from $\gamma$-rays of the same energy in a typical scintillator \cite{smith99}.

%
\section{Operational performance of ZEPLIN-II during the first science run}
Results are presented in this paper from the first 57 day underground, fully shielded, science run of ZEPLIN-II. Table~\ref{exposure_table} summarises the exposure cuts applied to this data run, illustrating the significant exposure reduction required due to the fiducial volume cuts discussed above. Periods during which the extraction field experienced fluctuations in applied voltage were excluded from this run, through removal of that entire day of data. Daily $\gamma$-ray calibrations were performed and routine maintenance on the Polycold cooling system also reduced the science exposure. Ultimately 225~kg$\times$days of data were included in the following analysis, from a live time of 31.2~days.

\subsection{Energy calibration, position and energy resolution and nuclear recoil scintillation efficiency}
\label{qf}
\begin{figure}[p]
  \centerline{\epsfig{file=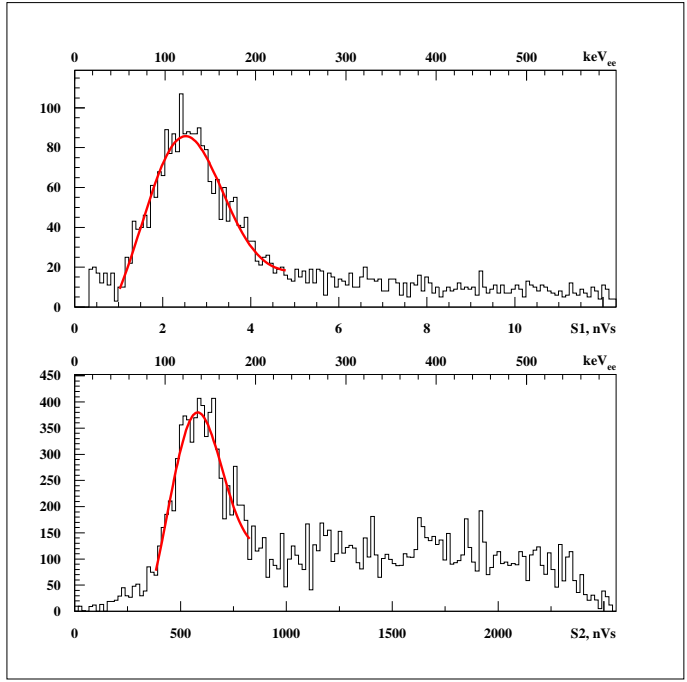,width=9cm}}
  \caption{Typical energy spectra for $^{57}$Co $\gamma$-ray calibrations, showing S1 spectrum (upper) and S2 spectrum (lower). The fits are double Gaussian fits which incorporate both the 122~keV and 136~keV lines in the $^{57}$Co $\gamma$-ray spectrum. The energy resolution of the detector is derived from the width of the S1 peak, coupled with calibration measurements at other line energies.}
  \label{co57_cal} 
\end{figure}

\begin{figure}[p]
  \centerline{\epsfig{file=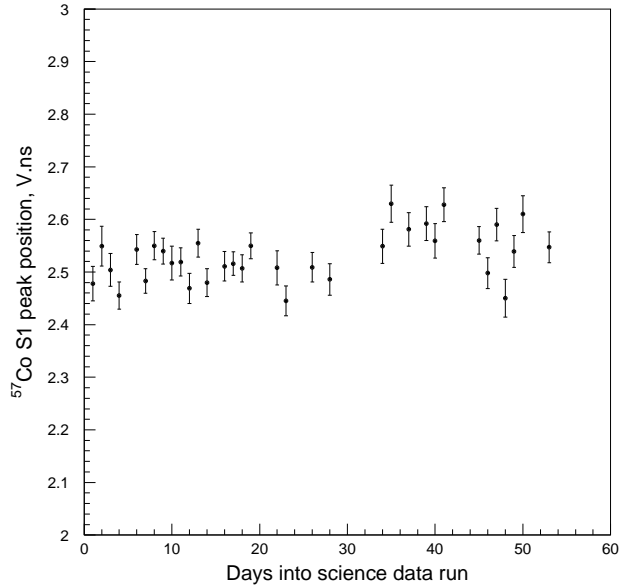,width=9cm}}
  \caption{Time evolution of the mean primary scintillation response to 122~keV $^{57}$Co $\gamma$-rays obtained from the regular calibration runs performed during the extended science run.}
  \label{S1_stability} 
\end{figure}

To calibrate the photomultiplier output in terms of electron recoil energy, a $^{57}$Co $\gamma$-ray source was used, placed between the detector vessel and the liquid scintillator veto by an automated source delivery mechanism.  The copper base was made thinner in various places to allow the 122~keV and 136~keV $\gamma$-rays to penetrate through to the bottom 1~cm layer of liquid xenon and make visible the combined photopeak, as shown in Fig.~\ref{co57_cal}.   This allowed a numerical value to be obtained for the photoelectron yield for the photomultiplier array and for the individual photomultipliers for the primary scintillation signal, the parameter used as a measure of the energy of an interaction.  The $^{57}$Co calibration was carried out daily.  The average photoelectron yield for the photomultipliers was 1.10$\pm$0.04 photoelectrons/keV with the electric drift field set to zero, and 0.55$\pm$0.02 photoelectrons/keV with the electric field at its operating value of 1~kV/cm.   The factor of two difference in light yield arises because the recombination component of the scintillation light is suppressed by the removal of charge by the electric field. The stability of the primary signal for these $^{57}$Co $\gamma$-ray interactions during this science run is shown in Fig.~\ref{S1_stability}.

Electron lifetime measurements gave an average figure of 112~$\mu$s. The observed light collection throughout the active volume was uniform to within 3\%, determined from the primary scintillation signals of $\alpha$-particle events which are uniformly distributed as a function of depth. Detailed light collection Monte Carlo simulations show that to achieve this uniformity an absorption length of $>$100~cm for the VUV photons is required.

\begin{figure}[p]
  \centerline{\epsfig{file=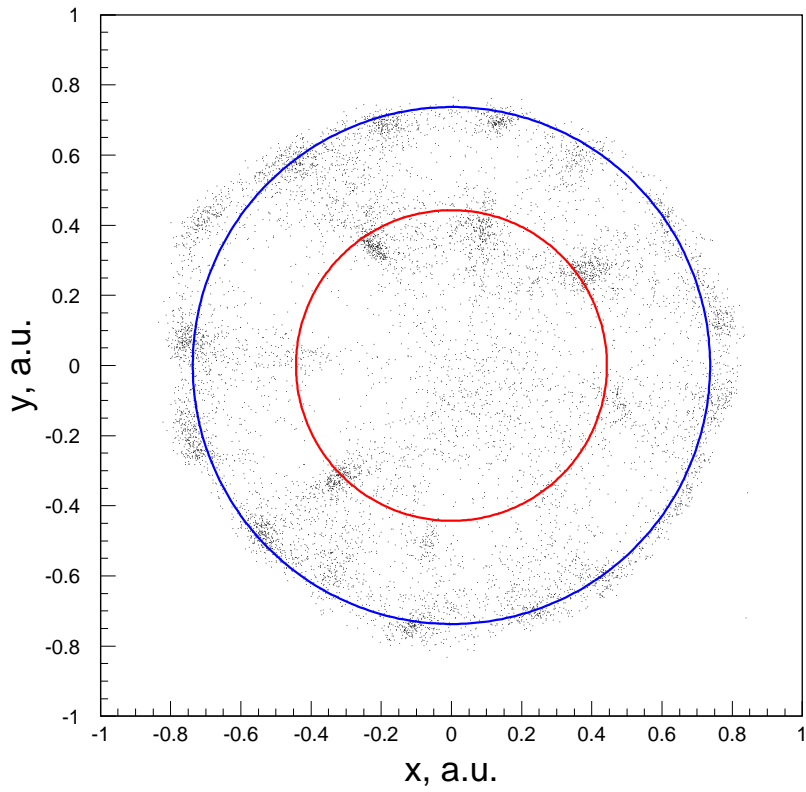,width=12cm}}
  \caption{Position reconstruction of 122/136~keV $^{57}$Co $\gamma$-rays, showing the two concentric rings of pits in the copper base plate of the target which allow the $\gamma$-rays to enter the active volume of xenon.}
  \label{posn_cal} 
\end{figure}

The $^{57}$Co $\gamma$-ray calibrations also provide the ability to calibrate the position reconstruction algorithm for interactions within ZEPLIN-II. The thinned regions within the base plate of the active volume are in the form of pits located on two concentric circles of radius 7.5~cm and 15~cm. Fig.~\ref{posn_cal} shows the reconstructed positions of $^{57}$Co $\gamma$-rays, using the secondary scintillation signals, in which the recessed pits are clearly visible. Note that this position reconstruction is performed near the bottom cathode of the detector, thereby at the extremity of the electron drift length where any lateral diffusion of the drifting charge cloud will be at a maximum.

The energy resolution, determined from the width of the $^{57}$Co 122/136~keV $\gamma$-ray peak and other calibration lines, was $\sigma_E$ =(1.80$\pm$0.04)$\times\sqrt{E}$ [keV], with E being the $\gamma$-ray energy in keV.   This has the effect of mixing the events between energy bins, which can at the final stage of analysis be accounted for by applying a compensating rebinning matrix to the energy-binned spectral terms, as shown in detail in \cite{zeplin1}.

The relative scintillation efficiency or Òquenching factorÓ for nuclear recoils, has now been measured in liquid xenon by several groups \cite{akimov02,chepel05,aprile05} giving an average value QF=0.19$\pm$0.02 \cite{chepel05} which remains constant with energy in the few 10's~keV nuclear recoil energy.  For this analysis a constant quenching factor was used, although Ref.~\cite{aprile05} may indicate some reduction at lower energies, where the nuclear recoil detection efficiency for ZEPLIN-II is low. This relative scintillation efficiency refers to the scintillation output relative to that from electron recoils at zero electric field.   When expressed relative to the field suppressed scintillation output from electron and nuclear recoils in a field of 1~kV/cm, the conversion factor between electron and nuclear recoil energy becomes $E_{nr} = E_{ee}/QF\times(f_e/f_n)$ where $f_e=0.50$ is the field induced suppression for electron recoil scintillation obtained from the $^{57}$Co calibration and  $f_n=0.93$\cite{aprile05} is that for nuclear recoils. Therefore $E_{nr} = E_{ee}$/0.36.

\subsection{Low energy neutron and $\gamma$-ray calibrations}
\label{calibrations}
Calibration of the ZEPLIN-II detector response to neutrons and  $\gamma$-rays was performed underground using neutron and  $\gamma$-ray sources. Neutrons (and  $\gamma$-rays) were provided by 0.1~GBq and 0.3~GBq AmBe sources manually located within the neutron shielding at a distance of $\sim$1~m from the active volume. A near uniform population of low energy Compton scattered  $\gamma$-rays was also provided as a comparison by a manually delivered 10~$\mu$Ci $^{60}$Co source, again inserted inside the detector shielding. These calibrations were performed at a significantly higher DAQ rate than the science run. Although this ensured minimal contamination of the calibration data sets by background, an increase in random coincidences between real events and events with primary scintillation only (arising from dead regions of the detector where there is a reverse drift field) was observed. This led to a uniform distribution of events in the S2/S1 parameter space, confirmed through study of events with unphysical drift times, ie those with an apparent location beyond the maximum drift distance of the active volume.

\begin{figure}[p]
  \centerline{\epsfig{file=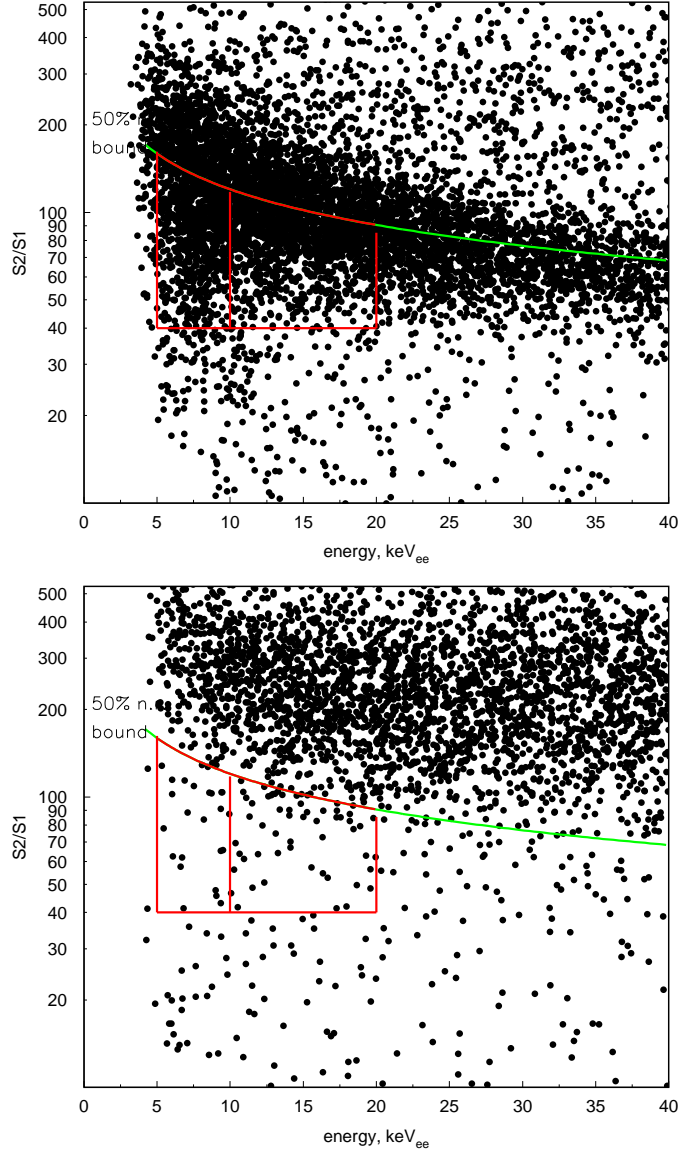,width=10cm}}
  \caption{Calibration using neutrons from an AmBe source (upper) and Compton scattered $^{60}$Co $\gamma$-rays (lower). These calibrations were performed at a high trigger rate, leading to a small, uniform, population of coincidental events distributed throughout the S2/S1 parameter space. These coincidentals are verified by comparison with events with unphysical drift times. Also shown are the S2/S1 boundary for 50\% nuclear recoil acceptance, and the acceptance window used in the dark matter analysis.}
  \label{ambeco60_cal} 
\end{figure}

As shown in Figs.~\ref{pulse} and \ref{n_pulse} the neutrons and  $\gamma$-ray interactions have differing values of the ratio of primary scintillation to secondary electroluminescence (S2/S1). This provides the discrimination power of the two-phase xenon technique, and is illustrated in Fig.~\ref{ambeco60_cal} where the two calibration populations are shown in the log(S2/S1) vs. S1 parameter slice, i.e. where the secondary electroluminescence signal is normalised to the event energy. The two event populations have differing centroids of S2/S1, which separates as the energy of the event interaction increases. The region that contributes dominantly to the dark matter limit lies below 25~keV$_{ee}$ for liquid xenon, due to the nuclear form factor, where the two distributions become broader and show some degree of overlap. At higher energies (outside the range of Fig.~\ref{ambeco60_cal}) a second population was observed, arising from inelastic neutron scattering from $^{129}$Xe.   Comparison of the AmBe neutron and $^{60}$Co $\gamma$-ray calibrations shows a discrimination power against $\gamma$-rays of 98.5\% for 50\% acceptance of nuclear recoils, between 5~keV$_{ee}$ and 20~keV$_{ee}$, the region of interest for dark matter searches. This value is applicable only to the operational charge collection field of 1~kV/cm used during this science run. Although not originally designed for charge drift field $\gg$1~kV/cm there is evidence that increasing this drift field will enhance charge extraction from electron recoils  \cite{voronova89}, thereby affording greater discrimination between of nuclear recoils, which will be explored in future runs.

\begin{figure}[p]
  \centerline{\epsfig{file=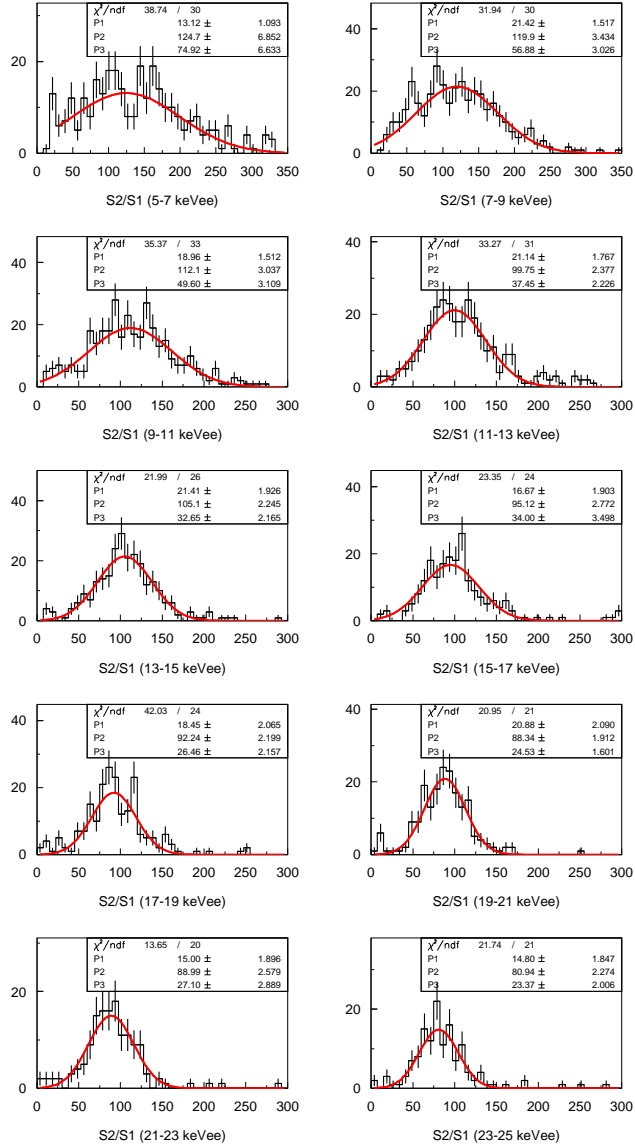,width=10cm}}
  \caption{Definition of the nuclear recoil acceptance window. Gaussian-fitting of the AmBe S2/S1 distributions binned in 2~keV$_{ee}$ intervals. A nuclear recoil acceptance window is then defined for the 5Ñ20~keV$_{ee}$ range.}
  \label{ambe_sliced} 
\end{figure}

The nuclear recoil acceptance region used for the WIMP searches was determined from the AmBe and $^{60}$Co calibrations. This acceptance window was defined between 5~keV$_{ee}$ and 20~keV$_{ee}$ in energy and from a baseline of S2/S1=40 up to an S2/S1 value which provides 50\% nuclear recoil acceptance. The definition of this value is shown in Fig.~\ref{ambe_sliced} where the differential AmBe neutron event distributions are plotted and fitted in (S2/S1) for various energy bands. Integrating the fitted neutron populations provides the fraction of the nuclear recoil population which lies to the left of any chosen value of (S2/S1). Validation of the positioning of this acceptance window by studying 10\% of the science data was undertaken to ensure consistency between the neutron and $\gamma$-ray calibrations and the science data. Following this comparison the acceptance window was frozen for subsequent analysis of the full science data-set.

Thus the aim of the experiment is to carry out extended science runs without calibration sources, to look for events in the nuclear recoil region, or to set confidence limits on their rate, and hence upper limits on the WIMP-nucleon cross-section. In practice, rather than zero observed events, there will be overlapping $\gamma$-ray events in this region and potentially residual neutrons or other background events, in which case the Feldman-Cousins limit \cite{feldman98} on that number can be used.  In addition, according to the number of background events in this Òacceptance regionÓ, one has the option of subdividing the region into several energy ranges and combining the separate limits for each. 

\subsection{Data stability and secondary signal corrections}
\label{corrections}
During the extended science run the stability of the detector was monitored extensively through a dedicated slow control system, including gas phase pressure, target temperatures, cooling system temperatures, photomultiplier trigger rates, DAQ trigger rate, liquid xenon purity, photomultiplier single photoelectron size and energy calibrations. As shown in $\S$\ref{qf} the response of the detector to the primary scintillation signal was uniform throughout the entirety of the science data run, which also illustrates the stability of the photomultiplier and DAQ systems, as also determined directly.

Due to a minor coolant leak within the Polycold circulation system, the efficiency of the cooling system was not uniform throughout the science run, being more efficient when the coolant reservoir was recharged. Although temperature was controlled on the liquefaction head, this variability in cooling power had the effect of varying the target gas pressure and environment temperature during the run. Accordingly, this varied the secondary electroluminescence photon production, directly by changing the electroluminescence gas pressure and indirectly by changing the xenon liquid level between the extraction grids, and hence the electroluminescence field and path length in the gas. In addition, during the extended run the electron lifetime within the xenon varied about the average of 112~$\mu$s, which affects the secondary signal size through attenuation of the charge cloud during drifting to the xenon liquid surface.

\begin{figure}[p]
  \centerline{\epsfig{file=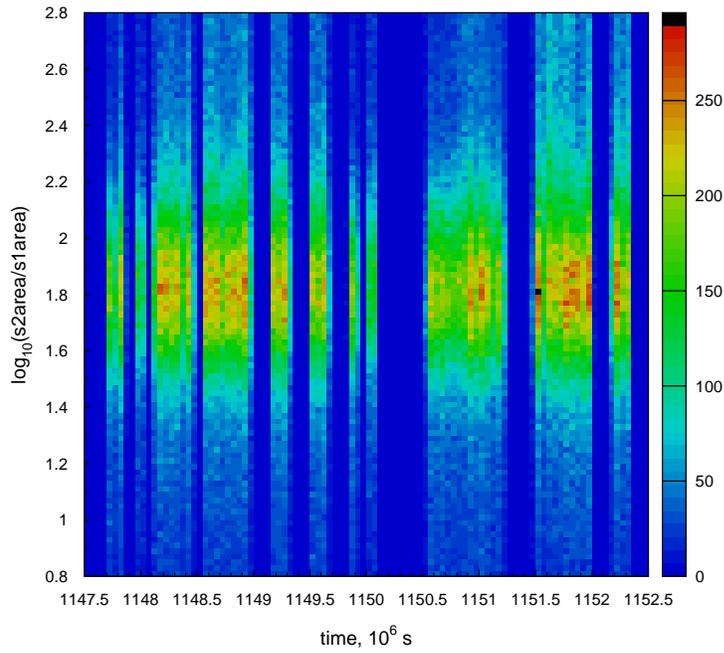,width=10cm}}
  \caption{Time evolution of S2/S1 distributions for nuclear recoil events from the cathode, following corrections to S2 for liquid xenon electron lifetime, electroluminescence gas pressure and temperature and residual surface charging. The stability of S2/S1 for this cathode nuclear recoil population, assumed to inject a constant charge distribution with time, is used as a normalisation for events within the active xenon volume.}
  \label{evol_cathode}
\end{figure}

To correct for these variations in the secondary signal size the science data and calibration charge yields were normalised on an event by event basis throughout the run length. To correct for the electron lifetime within the xenon bulk, the purity of the xenon was calculated every 4 hours by comparing the charge yield for nuclear recoils from the cathode against those from the extraction grid located under the liquid surface. The assumption during this normalisation is that the two populations have a constant charge yield distribution with time, verified as this technique gave an average lifetime of 112~$\mu$s, consistent with that calculated from internal  $\alpha$-particle and $\gamma$-ray events within the fiducial volume. The pressure effect on S2 size was measured directly in a dedicated run and corrected accordingly. To account for any residual S2 variability due to electroluminescence field variations and liquid surface charging (due to $<$100\% electron extraction) a final correction was applied by studying the S2/S1 ratio from nuclear recoils on the cathode. The stability of S2/S1 for this population, assumed to have a constant charge yield distribution with time, was used as a normalisation for events within the xenon bulk. Overall the S2/S1 corrections have a maximum variation in log space of $\pm$10\% from the mean, excluding the xenon purity correction. Fig.  \ref{evol_cathode} shows the time evolution of the nuclear recoil population on the cathode after all corrections have been applied, illustrating the stability of the corrected S2 signal.

\subsection{Event selection and detector response efficiencies}
\label{efficiencies}
The energy-dependent detector response function and event selection efficiency factor $ \eta(E)$, applied in $\S$\ref{get_the_limit}, relates the observed number of events in the fiducial volume to the actual number of interactions. This efficiency factor is a combination of several efficiency losses, including the hardware trigger, event selection cuts and the event search algorithm. For each event selection cut applied, detailed in Table \ref{efficiency_table}, the efficiency for nuclear recoils has been independently determined from source calibrations, dedicated data runs or simulations, as appropriate. The individual efficiencies for each selection cut are shown in Table \ref{efficiency_table}, illustrating that the main efficiency losses are due to the event trigger, DAQ dead-time and, at higher energies, DAQ saturation on large secondary pulses.

\begin{figure}[p]
  \centerline{\epsfig{file=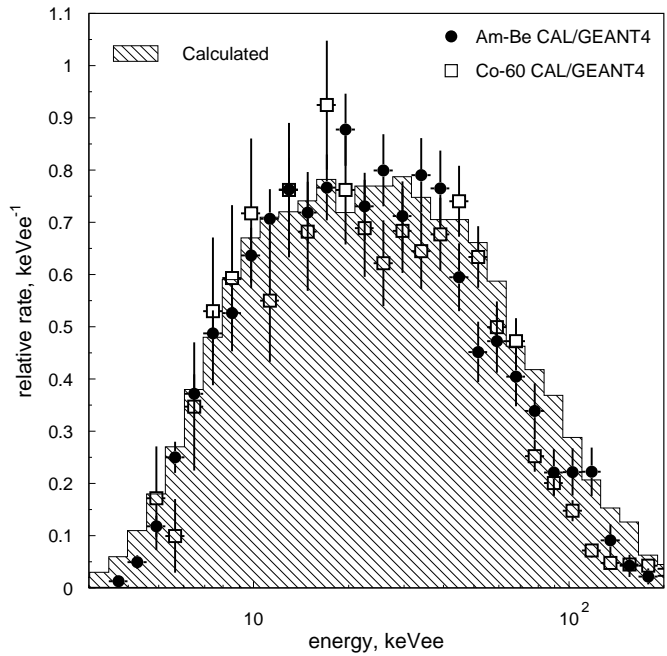,width=10cm}}
  \caption{Comparison of the overall nuclear-recoil detection efficiency, calculated as described in $\S$~\ref{efficiencies}, with the relative efficiencies obtained by dividing the actual AmBe and $^{60}$Co calibration spectra by the simulated energy dependencies obtained for single scatters in a detector with unity efficiency. Although the calibration data have been scaled to match the calculated efficiency (hatched region), the good agreement in spectral shape supports the calculations summarised in Table~\ref{efficiency_table}.}
  \label{eff_comparison}
\end{figure}

To verify the overall event selection and detector response efficiencies the combined efficiency of all factors in Table \ref{efficiency_table} is compared to the AmBe and $^{60}$Co $\gamma$-ray source calibrations. This comparison is shown in Fig.  \ref{eff_comparison}, where the event spectra for each calibration, normalised to the energy distribution derived from a GEANT4 \cite{geant4} simulation of single scattered interactions due to each source, is compared against the calculated overall detector response efficiency shown by the hatched histogram in the figure. Good agreement between the normalised event spectra and the calculated efficiency factor  $ \eta(E)$ is seen, especially in the relevant low energy region, verifying the combination of all individual factors.

\subsection{Differential energy spectrum of electron recoils in the science run}
\begin{figure}[p]
  \centerline{\epsfig{file=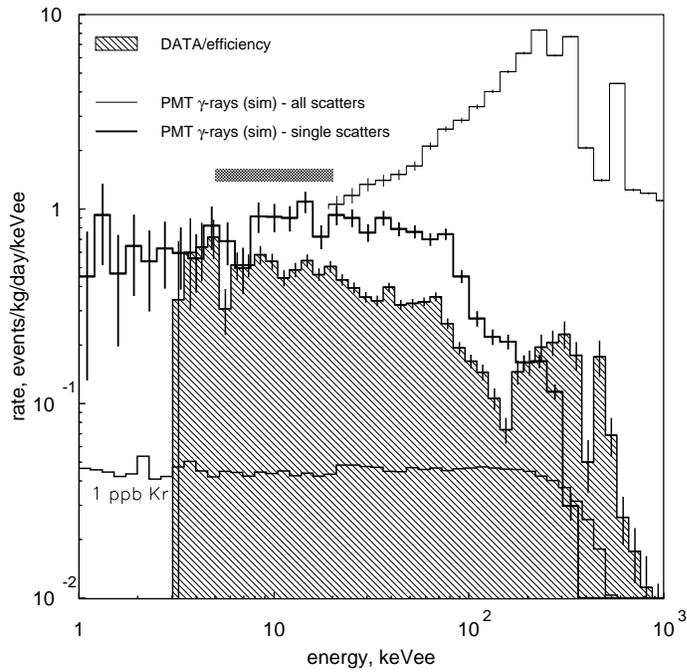,width=10cm}}
  \caption{The observed differential energy spectrum of electron recoils observed in the fiducial volume during the science run. The data, corrected for detector efficiencies, are shown as the hatched region. For comparison the expected event rate from a GEANT4 simulation of the photomultiplier $\gamma$-ray background is shown, for single and multiple scatters, showing good agreement in the region of interest. Also shown is the expected background from a nominal 1~ppb contamination of Kr which, when compared to the observed spectrum, limits the Kr contamination to below the 30-40~ppb level originally assumed from manufacturers specifications.}
  \label{bkdru}
\end{figure}

The differential energy spectrum of the electron-recoil background in the fiducial volume during the science run is shown in Fig.~\ref{bkdru}, taking into account the energy-dependent efficiencies. In the 5--20~keV$_{ee}$ energy range of interest (indicated by the shaded area) the background rate averages 0.5~evts/kg/day/keV$_{ee}$.

The figure also shows the result of GEANT4 simulations of the photomultiplier $\gamma$-ray background used hitherto to predict the instrument sensitivity. These considered the U/Th/$^{40}$K contaminants as indicated by the manufacturer. The simulated background is slightly higher than the measured values, but agreement is relatively good in the low-energy Compton region up to $\sim$100~keV$_{ee}$. Above this energy a microscopic model which takes into account the spatial extent of each interaction in the xenon is required, and this was not considered in this simulation.

The differential spectrum expected from $^{85}$Kr decay in the target is also shown in the figure, for a token contamination of 1~ppb Kr which can be easily scaled. The actual electron-recoil background seems incompatible with a contamination in 30--40 ppb range as considered, rather conservatively, in the instrument design stage.

In conclusion, although it is clear from the z-dependence of the event distribution (not shown) that not all background within the fiducial volume is due to the photomultiplier array, the observed background rate within the detector is close to that originally expected from simulations of the photomultiplier contaminants.

%
\section{Dark matter cross-section limit calculations}
\begin{figure}[p]
  \centerline{\epsfig{file=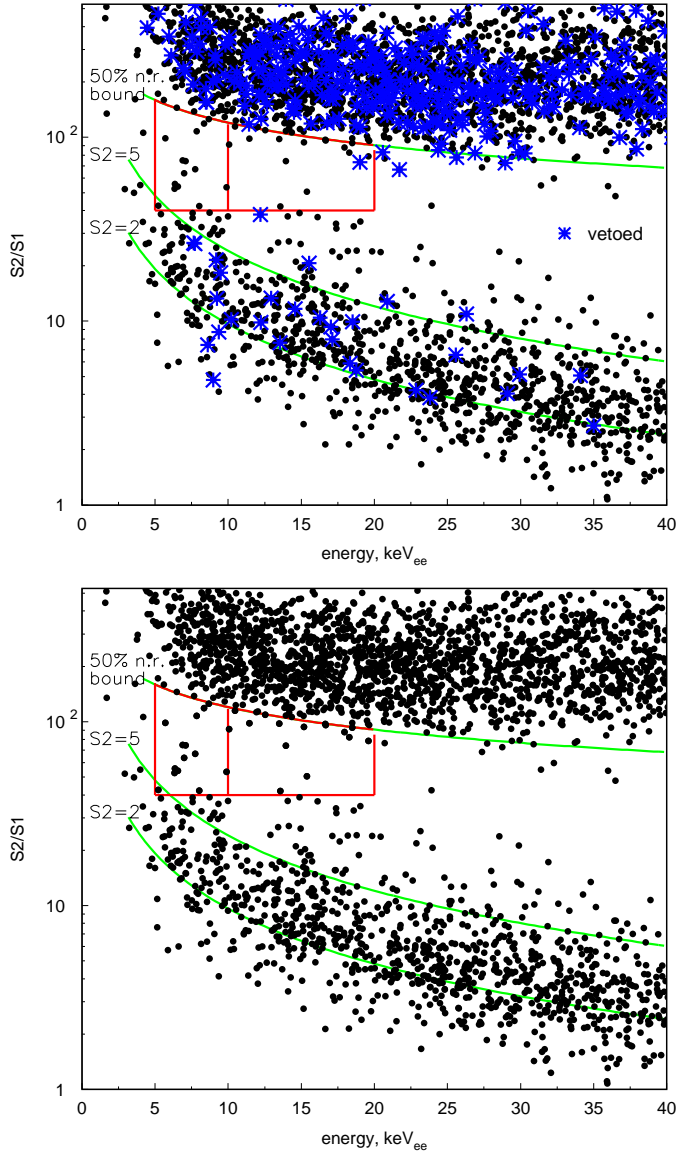,width=10cm}}
  \caption{Science data from the complete 225~kg$\times$day exposure of ZEPLIN-II, with secondary signal sizes corrected and normalised as described in $\S$\ref{corrections}. The data are shown in the S2/S1 vs energy space used for the neutron and $\gamma$-ray calibrations. The upper plot shows events that also have a signal recorded in the liquid scintillator veto, the lower plot has these events removed. The nuclear recoil acceptance window used for the dark matter analysis is shown, with the 50\% nuclear recoil acceptance boundary extended across the energy range. Also shown are two contours of constant S2, showing the radon progeny background events observed in the lower S2/S1 population have a fixed S2 distribution.}
  \label{background_data} 
\end{figure}

Following the normalisation of the S2 electroluminescence signal size for electron lifetime, xenon gas pressure and temperature and surface charging discussed in $\S$\ref{corrections}, the science data from the 225~kg$\times$days exposure run are plotted into the same log(S2/S1) vs energy parameter space as used for the nuclear and electron recoil calibrations. Fig.~\ref{background_data} shows the complete dataset, with the upper plot also indicating events where a signal was recorded in coincidence with the liquid scintillator veto. These vetoed events are removed in the lower plot, which is then used to derive the dark matter limits for this run. Also shown in Fig.~\ref{background_data} is the acceptance window defined from the calibrations described in $\S$\ref{calibrations}, between 5~keV$_{ee}$ and 20~keV$_{ee}$, and from an S2/S1 of 40 to a value of S2/S1 equivalent to 50\% nuclear recoil acceptance. In total 29 events are seen within this acceptance window, as detailed in Table~\ref{expectation_table}. These events are clearly dominated by the overlapping tail of the $\gamma$-ray distribution and small nuclear recoil background events from the PTFE walls which spill into the acceptance window along constant S2 contours, in spite of the radial cut.

\subsection{Event expectations within the nuclear recoil acceptance window}
\begin{figure}[p]
  \centerline{\epsfig{file=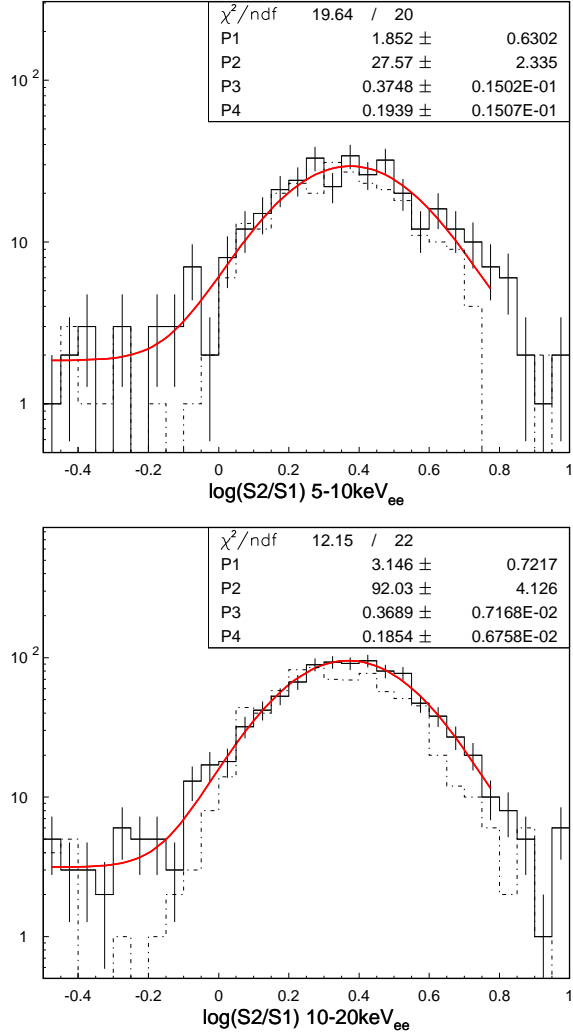,width=9cm}}
  \caption{Calculation of the expected $\gamma$-ray count in the acceptance window from the $^{60}$Co calibration. The differential event rate is shown for several energy slices as a function of log(S2/S1), normalised to the 50\% nuclear recoil acceptance value. A Gaussian+offset fit is made to the data, the offset accounting for the coincidental events arising from the high trigger rate used during this calibration. The expectation count for  $\gamma$-ray events in the science run is calculated by integrating the Gaussian, normalised to the overall event count in the science data, between the relevant values of log(S2/S1) and energy. The error on the expectation count is derived directly from the errors on the Gaussian fit. Also shown, dashed, are the science data distributions for these energy bands, illustrating the $\gamma$-ray nature of the background events in the science run}
  \label{gamma_exp} 
\end{figure}

The first, expected, population of events observed in the nuclear recoil acceptance region is from the overlapping tail of the $\gamma$-ray distribution. The expected number of events from this background is calculated from the $^{60}$Co $\gamma$-ray source calibration. Fig.~\ref{gamma_exp} shows the event rate for this calibration for the two relevant energy slices as a function of log(S2/S1). A Gaussian fit is made to the calibration data, with a uniform offset which accounts for coincidental events arising from the high trigger rate used during this calibration, as discussed in $\S$\ref{calibrations}. The expectation count for $\gamma$-ray events in the acceptance region for the science run is calculated by integrating the Gaussian, normalised to the overall event count in the science data, between the relevant values of log(S2/S1) for a given energy span. The error on the expectation count is derived directly from the errors on the Gaussian fit. Also shown in Fig.~\ref{gamma_exp} are the science data distributions for these energy bands, illustrating the $\gamma$-ray nature of the events in the science run. As a cross check on the $\gamma$-ray expectation, the expected number of events was also calculated directly from the science data itself. A Gaussian fit was made to the differential event rate in log(S2/S1) for a given energy span, including only values of S2/S1 above the acceptance region. The expectation was then derived by integrating the Gaussian between the relevant values of S2/S1. The predictions calculated from both techniques are shown in Table \ref{expectation_table}.

\begin{figure}[p]
  \centerline{\epsfig{file=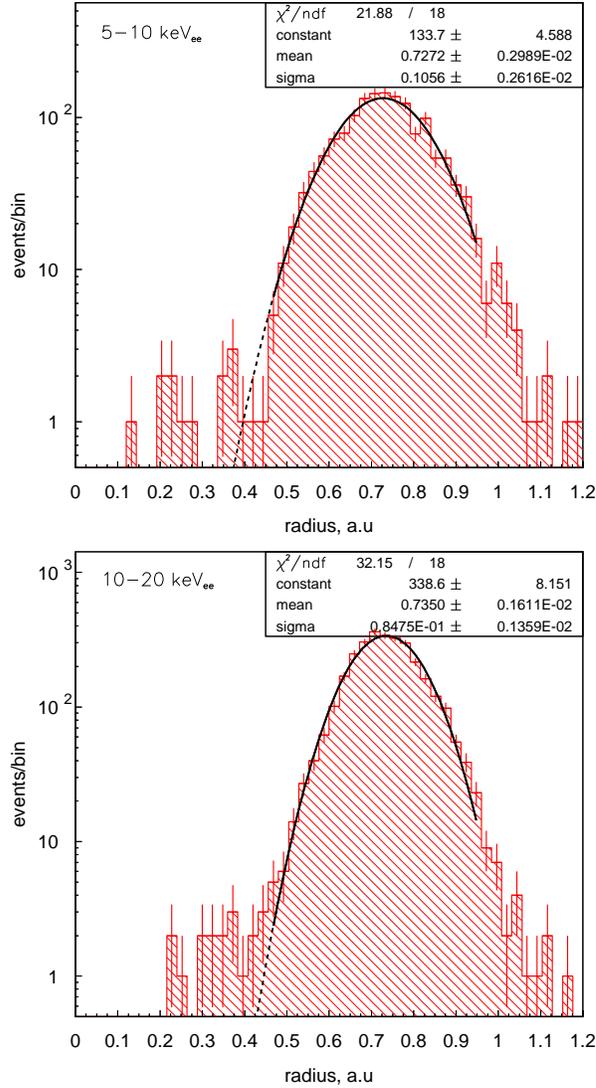,width=9cm}}
  \caption{Calculation of the expected background counts in the acceptance window due to the radon daughter ions plating on the PTFE side wall surfaces. The event number in the 5-10~keV$_{ee} (upper) and $10-20~keV$_{ee}$ (lower) acceptance regions are shown as a function of the reconstructed radius, showing the poor position reconstruction for this class of event, but confirming the location as the PTFE walls, which are at a radius of $\sim$0.75~a.u. Events at a radius $<$0.46~a.u. are those events observed in the acceptance window shown in Fig.~\ref{background_data}. A Gaussian fit is performed outside the radial cut used to define the fiducial volume and extrapolated into the fiducial volume to provide the expectation count. The error on this is derived directly from the errors on the Gaussian fit.}
  \label{bananas} 
\end{figure}

The second, unexpected, population of events which encroach on the acceptance window are seen to be nuclear recoil events of constant secondary size of $\sim$10 electrons. These are located on the PTFE walls of the active volume, but due to their small S2 signal have a poor position reconstruction accuracy, which results in a small fraction of these events being wrongly placed within the fiducial volume. These events are derived from radon nuclei decaying within the active volume, originally emitted from the SAES getters, which migrate to the PTFE walls and electrodes when positively charged following $\beta$-decay. Subsequent $\alpha$-decays  along the Rn-chain cause recoiling nuclei to enter the liquid xenon volume. Due to the proximity of these recoils to the PTFE walls there is incomplete charge extraction, or charge stripping as the electron cloud is drifted to the xenon surface, giving a poor S2 yield. The variation of the number of events within a given region of (S2/S1)-energy parameter space as a function of reconstructed radius is used to define the effective event location error for the nuclear recoils from the radon progeny induced wall events. Fig. \ref{bananas} shows this radial dependence for the 10-20~keV$_{ee}$ acceptance window. The expectation for these wall events is defined by the number of events where the event location error places a wall event (of the requisite primary energy) inside the fiducial volume. A Gaussian fit is performed to the distribution beyond the radial cut used to define the fiducial volume. This Gaussian is then extrapolated into the fiducial cut region and integrated to provide the expectation value, shown in Table \ref{expectation_table}.

\subsection{Nuclear recoil and WIMP-nucleon cross-section limits}
\label{get_the_limit}
From the observed and expected event count given in Table \ref{expectation_table} a 90\% confidence upper limit to the number of nuclear recoil events observed within the defined acceptance window may be derived from the Feldman-Cousins limit \cite{feldman98}. For the combined energy span of 5-20~keV$_{ee}$, where 29 events are seen and 28.6$\pm$4.3 are expected, this yields a 90\% c.l. of 10.4 nuclear recoil events within the 50\% nuclear recoil acceptance window in 225~kg$\times$days of exposure, using the mean expectation value, or an upper limit of 0.092~events/kg/day in total between 5 and 20~keV$_{ee}$. The TFeldmanCousins class within the ROOT analysis framework \cite{root} was used to extend the Feldman-Cousins tables in \cite{feldman98} into the relevant regime for this analysis. Although there is no uniquely accepted approach to determine the impact of the error on the expectation value, a Bayesian approach may increase this limit by no more than 20\%.

The event rate limit calculated from the mean expectation count is now compared with a theoretical dark matter spectrum in order to estimate an overall limit on the dark matter event rate, and hence a cross-section limit.  For a flux of particles of mass $M_D$~GeV incident on a nucleus of atomic number A, producing a nuclear recoil energy $E_R$~keV, the differential event rate R(events/kg/d) is given by Ref. \cite{lewin87}
\begin{equation}
\frac{dR}{dE_R}  =   \left(\frac{c_1R_0}{E_0r} \right)  \exp\left(\frac{-c_2E_R}{E_0r} \right)  F^2(E_R,A) ,
  \label{diff_rate}
\end{equation}
where  $E_0 = 0.5\times10^{-6}M_D (v0/c)^2$~[keV], $v_0$ = 220~km/s,  $r = 4M_DM_T/(M_D+M_T)^2$, $M_T$ = 0.932A, and $F^2$ is a nuclear form factor correction (discussed in Ref. \cite{lewin87}).  For a detector at rest with respect to an isotropic Maxwellian dark matter flux, $c_1 = c_2 = 1$, while motion through the Galaxy gives average fitted values $c_1$ = 0.75, $c_2$ = 0.56, with a small annual modulation tabulated in  \cite{lewin87}.  $R_0$ [events/kg/d] is defined as the total rate for a stationary Earth, and is related to the total nuclear cross-section $\sigma_A$~[pb] by
\begin{equation}
\frac{R_0}{r}   =  \frac{D\sigma_A }{ \mu_A^2} ,
  \label{total_rate}
\end{equation}
where $\mu_A = M_DM_T/(M_D+M_T)$ is the reduced mass of the colliding particles and D is a numerical factor equal to 94.3  \cite{lewin87} for an assumed dark matter density 0.3~GeV/cm$^3$.  $E_R$ is related to the experimentally observed electron equivalent energy $E_e$  by $E_R  = E_e /f_{Xe}$   and, as discussed in $\S$\ref{qf}, we take $f_{Xe}$ = 0.19 as the zero field value, corresponding to 0.36 when defined relative to the electron equivalent energy in a field of 1~kV/cm, as explained in $\S$\ref{qf}. 
It is customary to express the final limits in terms of the equivalent WIMP-nucleon cross-section $\sigma_{W-N}$~[pb], which is related to $\sigma_A$ , in the case of a spin-independent nuclear interaction, by
\begin{equation}
 \sigma_{W-N} =  \left(\frac{\mu_1}{\mu_A}\right)^2 \left(\frac{1}{A}\right)^2 \sigma_A(pb) ,
  \label{cross_sect}
\end{equation}
where $\mu_1$ ($\sim$0.925~GeV)  is the reduced mass for $A$ = 1. Hence, from (\ref{total_rate})		
\begin{equation}
 \sigma_{W-N}  \sim  9.1\times10^{-3}  \left(\frac{1}{A}\right)^2  \left(\frac{R_0 }{r }\right) ,
  \label{cross_sect2}
\end{equation}

We also need to include the energy-dependent experimental efficiency factor $ \eta(E) \leqslant 1 $ discussed in $\S$\ref{efficiencies}, translated into nuclear recoil energy $ \eta(E_R)$ and defined by 
\begin{equation}
\left[ \frac{dR}{dE_R}\right]_{actual} = \left[ \frac{1}{\eta(E_R)}\right] \left[ \frac{dR}{dE_R}\right] _{observed} .
 \label{eff}
\end{equation}
Thus the WIMP-nucleon cross-section limit setting procedure is 
\begin{enumerate}
\item{Apply an energy resolution correction as described in greater detail in a previous paper \cite{zeplin1}, by numerically applying the resolution rebinning matrix to the vector of binned spectral terms given by the right hand side of (\ref{diff_rate})}
\item{Set $R_0  = 1$, multiply the right hand side of (\ref{diff_rate}) by $\eta(E_R)$,  and numerically integrate (\ref{diff_rate}) over the energy span adopted for the observed rate limit, corresponding to an electron-equivalent range 5 - 20~keV. }
\item{Divide the observed rate limit by this integral to obtain the corresponding limit for $R_0$  .}
\item{Use (\ref{eff}) to convert to $\sigma_{W-N}$ (pb), repeating the process for each value of dark matter particle mass  $M_D$ .}
\end{enumerate}

\begin{figure}[p]
  \centerline{\epsfig{file=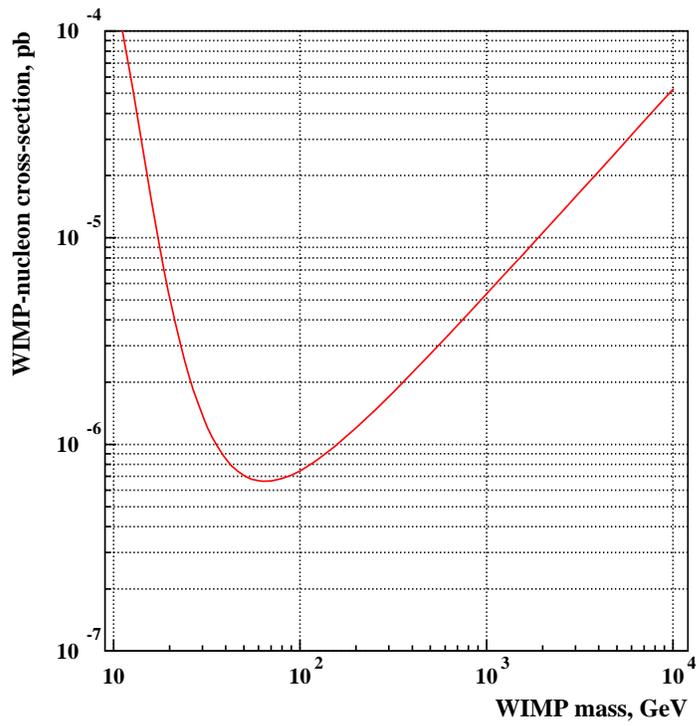,width=10cm}}
  \caption{The 90\% c.l. upper limit on the cross-section of WIMP-nucleon spin-independent interactions. The minimum of the cross-section limit lies at $6.6\times10^{-7}$~pb at a WIMP mass of 65~GeV}
  \label{final_limit} 
\end{figure}

Using this procedure, Fig. \ref{final_limit} shows the 90\% confidence upper limit to the spin-independent WIMP-nucleon cross-section derived from the 10.4 event upper limit to the nuclear recoil events in the defined acceptance window from the first 225~kg$\times$days exposure run of ZEPLIN-II. The minimum of this limit lies at $6.6\times10^{-7}$~pb at a WIMP mass of 65~GeV. 

%
\section{Conclusions}
First results are presented from a 31~day live run of ZEPLIN-II, a 31~kg two-phase xenon detector developed to observe nuclear recoils from hypothetical weakly interacting massive dark matter particles. The total exposure of this run following the application of fiducial and stability cuts was 225~kg$\times$days. Discrimination between nuclear recoils and background electron recoils is demonstrated using a AmBe neutron and $^{60}$Co $\gamma$-ray sources, allowing the definition of a nuclear recoil acceptance window of 50\% nuclear recoil acceptance between 5~keV$_{ee}$ and 20~keV$_{ee}$. This acceptance region registered 29 events in the science run, with a summed expectation of 28.6$\pm$4.3 $\gamma$-ray and radon progeny induced background events giving a 90\% c.l. upper limit to the number of nuclear recoils of 10.4 events in this acceptance window. This converts to a WIMP-nucleon spin-independent cross-section with a minimum of  $6.6\times10^{-7}$~pb following the inclusion of all detector and interaction efficiencies. 

A second extended science run of ZEPLIN-II is currently in preparation, in which the radon emission from the SAES getters will be eliminated, thereby removing the PTFE wall background population. The removal of this population, coupled with an expected increase in the fiducial active volume possible due to their removal, gives a projected sensitivity of $2\times10^{-7}$~pb for this second run, assuming a similar live time of $\sim$30 days, or  $1\times10^{-7}$~pb from an extension of the runtime to five months.

\section{Acknowledgements}
This work has been funded by the UK Particle Physics And Astronomy Research Council (PPARC), the US Department of Energy (grant number DE-FG03-91ER40662) and the US National Science Foundation (grant number PHY-0139065).  We acknowledge support from the Central Laboratories for the Research Councils (CCLRC), the Engineering and Physical Sciences Research Council (EPSRC), the ILIAS integrating activity (Contract R113-CT-2004-506222), the INTAS programme (grant number 04-78-6744) and the Research Corporation (grant number RA0350). We also acknowledge support from Funda‹o para a Cincia e Tecnologia (project POCI/FP/FNU/63446/2005), the Marie Curie International Reintegration Grant (grant number FP6-006651) and a PPARC PIPPS award (grant PP/D000742/1). We would like to gratefully acknowledge the strong support of Cleveland Potash Ltd., the owners of the Boulby mine, and J. Mulholland and L. Yeoman, the underground facility staff. We acknowledge valuable advice from Professor R. Cousins on the application and extension of the Feldman-Cousins tables.

%


\section{Tables and Figures}

\begin{sidewaystable}
\centering
\caption{Exposure summary for the first extended underground science run of ZEPLIN-II}
\label{exposure_table}
\vspace{5mm}
\begin{tabular}{l|l|l}
   & Exposure & Fiducial/operational cuts applied\\
\hline
Calendar runtime & 57 days & Overall length of run following operational parameter settings\\
(31 kg target mass)	& (1767 kg$\times$days) &\\
\hline
Science data run & 44.2 days & Science data exposure, removing calibrations and maintenance periods\\
(31 kg target mass)	 & (1370 kg$\times$days) &\\
\hline
Stable operation & 31.2 days & Removing days experiencing E-field instability\\
(31 kg target mass) & (967 kg$\times$days)	& \\
\hline
Fiducial cuts (drift time) & 31.2 days & Fiducial cut in z to remove near-grid events\\
(26 kg target mass) & (811 kg$\times$days) &\\
\hline
Fiducial cuts (radial)  & 31.2 days & Fiducial cut in x,y to remove side wall events\\
(7.2 kg target mass) & (225 kg$\times$days)	 & \\
\end{tabular}
\end{sidewaystable}

\begin{sidewaystable}
\centering
\caption{Event selection efficiencies}
\label{efficiency_table}
\vspace{5mm}
\begin{tabular}{r|l|l}
Selection cut & Efficiency & Description\\
\hline
S2 Cut-0 & $\approx$100\% (exp) & Requirement that a WIMP-like event has one and only one primary and secondary\\
S2 Cut-1 & f(E): 100\% $>$10~keV & Selection of S2 candidates with area$>$1~Vns \\
 & & (smaller pulses due to extraneous single electron extraction are ignored)\\
S2 Cut-2 & 90.2\% & Removal of events by S2 pulse shape cut (photon mean arrival time)\\
S2 Cut-3 & $\approx$100\% & Removal of events with non-physical S2 arrival times relative to trigger\\
S2 Cut-4 & $\approx$100\% & Removal of events with multiple S2 candidates (multiple scattering)\\
S1 Cut-1 & f(E): 100\% & Selection of S1 candidates with $\ge$3-fold coincidence at 2/5~photoelectron amplitude\\
 & (5~keV:43\% 10~keV:92\%) & \\
S1 Cut-2 & $\approx$100\% & Removal of events with non-physical drift times relative to S2\\
S1 Cut-3 & $\approx$100\% & Removal of events by S1 pulse shape cut (photon arrival time distribution)\\
S1 Cut-4 & 98.7\% & Removal of events with multiple S1 candidates\\
S1 Cut-5 & 99.7\% & Tagging of $<$3-fold S1 signals with cathode drift time (event removed by S1-4)\\
DAQ Cut-1 & f(E): 100\% $<$30~keV  & Digitiser saturation cut\\
DAQ Cut-2 & 90\% & DAQ dead-time correction for science run (trigger rate dependent)\\
DAQ Cut-3 & 99.2\% & Coincidental events in veto (trigger rate dependent)\\
DAQ Cut-4 & 99.7\% & Requirement that a valid S1 or S2 trigger the DAQ\\
\end{tabular}
\end{sidewaystable}

\begin{table}
\centering
\caption{Overall expectation values in the nuclear recoil acceptance window compared to observed counts}
\label{expectation_table}
\vspace{5mm}
\begin{tabular}{c|c|c|c|c|c}
Energy Range & Observed  & $\gamma$-ray ($^{60}$Co) (1) & $\gamma$-ray (data) & Rn-initiated (2) & Total (1+2)\\
\hline
5-10~keV$_{ee}$ & 14  & 4.2$\pm$2.4 & 5.6$\pm$4.6  & 10.2$\pm$2.2 & 14.4$\pm$3.3\\
10-20~keV$_{ee}$ & 15 & 11.9$\pm$2.7 &  13.0$\pm$6.0 & 2.3$\pm$0.5 & 14.2$\pm$2.7\\
\end{tabular}
\end{table}

\clearpage

\end{document}